\documentstyle[11pt,newpasp,twoside,epsf]{article}
\def\edcomment#1{\iffalse\marginpar{\raggedright\sl#1\/}\else\relax\fi}
\reversemarginpar

\begin{document}

 \title{The Probability Distribution of the Double Neutron 
Star Coalescence Rate and Predictions for More Detections }
 \author{Vassiliki Kalogera, Chunglee Kim}
 \affil{Northwestern U., Dept. of Physics \& Astronomy, 2145 Sheridan Rd.,
Evanston, IL 60208, USA}
 \author{and Duncan R. Lorimer}
 \affil{U. of Manchester, Jodrell Bank Observatory, Macclesfield,
Cheshire, SK11 9DL, UK}

 \begin{abstract}
 We present an analysis method that allows us to estimate the
Galactic formation of radio pulsar populations based on their
observed properties and our understanding of survey selection
effects. More importantly, this method allows us to assign a
statistical significance to such rate estimates and calculate the
allowed ranges of values at various confidence levels. Here, we apply
the method to the question of the double neutron star (NS--NS)
coalescence rate using the current observed sample, and we find
calculate the most likely value for the total Galactic coalescence
rate to lie in the range $3-22$ Myr$^{-1}$, for different pulsar
population models.  The corresponding range of expected detection
rates of NS--NS inspiral are $(1-9) \times 10^{-3}$ yr$^{-1}$ for
the initial LIGO, and $6-50$ yr$^{-1}$ for the advanced LIGO. Based
on this newly developed statistical method, we also calculate the
probability distribution for the expected number of pulsars that
could be observed by the Parkes Multibeam survey, when acceleration
searches will alleviate the effects of Doppler smearing due to
orbital motions.  We suggest that the Parkes survey will probably
detect $1-2$ new binary pulsars like PSRs~B1913+16 and/or B1534+12.
 \end{abstract}

\section{Introduction}
 
The detection of the double neutron star (NS--NS) prototype PSR~B1913+16
as a binary pulsar (Hulse \& Taylor 1975) and its orbital decay due to
emission of gravitational waves have inspired a number of quantitative
estimates of the coalescence rate, $\cal{R}$, 
of NS--NS binaries (Clark et al.\ 1979;
Narayan et al.\ 1991; Phinney 1991; Curran \& Lorimer 1995). Significant
interest derives from their importance as gravitational-wave sources for
the upcoming ground-based laser interferometers (such as LIGO).

We present a newly developed statistical analysis that allows the
calculation of statistical confidence levels associated with rate
estimates. The method can be applied to any radio pulsar population. Here,
we consider PSR~B1913+16 (Hulse \& Taylor 1975) and PSR~B1534+12
(Wolszczan 1991). For different assumed distributions of pulsar properties
(luminosities, Galactic positions), we derive the probability distribution
function of the total Galactic coalescence rate weighted by the two
observed binary systems. The method involves the simulation of selection
effects inherent in all relevant radio pulsar surveys and a Bayesian
statistical analysis for the probability distribution of ${\cal R}$. The
small-number bias and the effect of the faint-end of the luminosity
function, previously identified as the main sources of uncertainty in rate
estimates (Kalogera et al.\ 2001) are {\it implicitly included} in this
analysis. We extrapolate the Galactic rate to cover the detection volume
of LIGO and estimate the most likely detection rates of NS--NS inspiral
events for the initial and advanced LIGO. Details of this work are given
in Kim et al.\ (2003; hereafter KKL).

In the second part of this paper, we modify our statistical method in a
way that allows us to calculate the probability distribution for the
number of pulsars that could be detected by the Parkes Multibeam survey
(hereafter PMB survey; Lyne et al.\ 2000; Manchester et al.\ 2001) PMB,
when the effects of Doppler smearing due to orbital motions are corrected
with acceleration searches.

\section{Pulsar Survey Selection Effects} 

 For a model pulsar population with a given spatial and luminosity
distribution, we determine the fraction of the total population which are
actually {\it detectable} by current large-scale pulsar surveys.  In order
to do this, we calculate the effective signal-to-noise ratio for each
model pulsar in each survey, and compare this with the corresponding
detection threshold.  Only those pulsars which are nominally above the
threshold count as being detectable. After performing this process on the
entire model pulsar population of size $N_{\rm tot}$, we are left with a
sample of $N_{\rm obs}$ pulsars that are nominally detectable by the
surveys.  By repeating this process many times, we can determine the
probability distribution of $N_{\rm obs}$, which we then use to constrain
the population and coalescence rate of NS--NS binaries. More details are
given in \S\,2 of Lorimer et al.\ (1993) and in KKL.

Here we discuss in a some detail the the Doppler smearing effect which is
significant for recent surveys with relatively long exposures. For binary
pulsars, we need to take account of the reduction in signal-to-noise ratio
due to the Doppler shift in period during an observation.  For
observations of NS--NS binaries, where the orbital periods are of the
order of 10 hours or less, the apparent pulse period can change
significantly during a search observation causing the received power to be
spread over a number of frequency bins in the Fourier domain.  As all the
surveys considered in this analysis search for periodicities in the
amplitude spectrum of the Fourier transform of the time series, a signal
spread over several bins can result in a loss of signal-to-noise ratio
$\sigma$. To take account of this effect in our survey simulations, we
need to multiply the apparent flux density of each model pulsar by a
``degradation factor'', $F=\sigma_{\rm binary}/\sigma_{\rm control}$.
Significant degradation occurs, when $F \ll 1$.  Using an analysis method
described in Camilo et al.\ (2000), we calculate the degradation factor
for the two pulsars we consider in this work.  As expected, we find that
surveys with the longest integration times are most affected by Doppler
smearing. For the PMB survey, which has an integration time of 35 min,
mean values of $F$ are 0.7 and 0.3 for PSR~B1913+16 and PSR~B1534+12
respectively. The greater degradation for PSR~B1534+12 is
due to its mildly eccentric orbit ($e \sim 0.3$ versus 0.6 for
PSR~B1913+16)  which results in a much more persistent change in apparent
pulse period when averaged over the entire orbit.  
In order to improve on the sensitivity to binary
pulsars, the PMB survey data are now being reprocessed using various
algorithms designed to account for binary motion during the integration
time (Faulkner et al.; these proceedings). For the Jodrell Bank
and Swinburne surveys (Nicastro et al.\ 1995; Edwards et al.\ 2001), which
both have integration times of order 5 min, we find $F \sim 0.9$ for both
systems.  For all other surveys, which have significantly shorter
integration times, no significant degradation is seen, and we take $F=1$.

 \section{Probability Distribution of Double Neutron Star Coalescence 
Rates }

 \subsection{Statistical Method}

 As already mentioned, we generate large numbers of
``observed'' pulsar samples by modeling the survey selection
effects and applying them to model populations of PSR~B1913+16--like
and PSR~B1534+12--like pulsars, separately. For a fixed value of $N_{\rm
tot}$, we use these ``observed'' samples to calculate their distribution, 
which we find to be very well described by a Poisson distribution:
 \begin{equation}
 P( N_{\rm obs}; \lambda )~=~\frac{\lambda^{N_{\rm
obs}}~e^{-\lambda}}{N_{\rm obs}!} \, ,
 \end{equation} 
 where $\lambda\equiv <N_{\rm obs}>$. With our Monte Carlo simulations 
we calculate $\lambda$ and find it to linearly correlate with $N_{\rm 
tot}$ (for values in the range $10-10^{4}$):
 \begin{equation}
 \label{eq:lamb}
 \lambda = \alpha N_{\rm tot},
 \end{equation}
 where $\alpha$ is a constant that depends on the properties (space and
luminosity distributions and pulse period and width) of the Galactic
pulsar population.

For a given $N_{\rm tot}$, we calculate the rate using estimates of the
associated pulsar beaming correction factor $f_{\rm b}$ and lifetime
$\tau_{\rm life}$: ${\cal R}~=~\frac{N_{\rm tot}}{\tau_{\rm life}} f_{\rm
b}$. We adopt values discussed in Kalogera et al.\ (2001): 5.72 and $3.65
\times 10^{8}$\,yr for PSR~B1913+16, and 6.45 and $2.9\times 10^{9}$\,yr
for PSR~B1534+12.

Using Bayes' theorem and the best-fit Poisson distributions, we can
calculate the probability distribution of the total number $N_{\rm tot}$
of pulsars in the Galaxy. Further, using estimates of the associated
pulsar beaming correction factor $f_{\rm b}$ and lifetime $\tau_{\rm
life}$, we can calculate the distribution function of pulsar rates: ${\cal
R}~=~\frac{N_{\rm tot}}{\tau_{\rm life}} f_{\rm b}$. We adopt values
discussed in Kalogera et al.\ (2001): 5.72 and $3.65 \times 10^{8}$\,yr
for PSR~B1913+16, and 6.45 and $2.9\times 10^{9}$\,yr for PSR~B1534+12.
The probability functions of coalescence rates of pulsars similar to each
of the observed one are then given by:
 \begin{equation}
 P({\cal R}) = \Bigl({\frac{\alpha \tau_{\rm life}}{f_{\rm b}}} \Bigr)^{2}
{\cal R}~ e^{-\bigl({\frac{\alpha \tau_{\rm life}}{f_{\rm b}}} \bigr)
{\cal R}}.
 \end{equation} 
 We use appropriate variable transformations to then calculate the total 
rate probability distribution:  
 \begin{equation}
P({\cal R}_{\rm tot}) = {\Bigl( {AB \over B-A}\Bigr)^{2}} \Bigl[ {{\cal
R}_{\rm tot}} {\bigl({ e^{-A{\cal R}_{\rm tot}} + e^{-B{\cal R}_{\rm tot}}
}\bigr)} - {\Bigl( {2 \over B-A} \Bigr)} {\bigl({ e^{-A{\cal R}_{\rm tot}}
- e^{-B{\cal R}_{\rm tot}} }\bigr)} \Bigr], 
 \end{equation} 
 where A and B are defined as follows:
 \begin{equation}
 A \equiv {\biggl({ \alpha \tau_{\rm life} \over f_{\rm b} }\biggr)}_{\rm
1913}~~{\rm and }~~ B \equiv {\biggl({ \alpha \tau_{\rm life} \over f_{\rm
b}}\biggr)}_{\rm 1534}.
 \end{equation} 
 Having calculated the probability distribution of the Galactic
coalescence rate, we can take one step further and also calculate ranges
of values for the rate ${\cal R}_{\rm tot}$ at various confidence levels
(CL). The lower (${\cal R}_{\rm a}$) and upper (${\cal R}_{\rm b}$)  
limits to these ranges are determined by the following conditions:
 \begin{equation}
 \int_{{\cal R}_{\rm a}}^{{\cal R}_{\rm b}} P({\cal R}_{\rm tot}) d{\cal
R}_{\rm tot} = {\rm CL}~~{\rm and}~~ P({\cal R}_{\rm a})=P({\cal R}_{\rm
b}).
 \end{equation} 
 
 Finally, we can calculate the detection rate for LIGO,
${\cal R}_{\rm det}$, defined by
 \begin{equation}
 {\cal R}_{\rm det} = \epsilon {\cal R}_{\rm tot} V_{\rm det},
 \end{equation} 
 where $\epsilon$ is the scaling factor (based on the blue luminosity
density of the nearby universe) derived to be $\simeq 10^{-2}$\,Mpc$^{-3}$
(for details see Kalogera et al.\ 2001).  $V_{\rm det}$ is the detection
volume defined as a sphere with a radius equals to the maximum detection
distance D$_{\rm max}$ for the initial ($\simeq 20$\,Mpc) and advanced
LIGO ($\simeq 350$\,Mpc; Finn 2001).

\subsection{Results} 

 We have chosen one of our pulsar population models to be our {\em
reference} model based on the results presented
by Cordes \& Chernoff (1997). For this model, we find the most
likely value of $N_{\rm tot}$, to be $\simeq 390$ pulsars for the
``PSR~B1913+16-like'' population, and $\simeq 350$ pulsars for the
``PSR~B1534+12-like'' population. Using eq.\ (4--5), we evaluate the total
Galactic coalescence rate of NS--NS binaries for this reference case.
The most likely value of the coalescence rate is ${\cal R}_{\rm
peak}$$\simeq 8$\,Myr$^{-1}$ and the ranges at different statistical
confidence levels are: $\sim 3-20$\ Myr$^{-1}$ at 68\%, $\sim 1-30$\
Myr$^{-1}$ at 95\%, and $\sim 0.7-40$\ Myr$^{-1}$ at 99\%.  Also, the most
likely values of detection rates, which correspond to ${\cal R}_{\rm
peak}$ are $\sim 3\times 10^{-3}$\,yr$^{-1}$ and $\sim 18$\,yr$^{-1}$, for
the initial and advanced LIGO.

%%% figure1
    \begin{figure}[t]
    \vbox to 8.0cm {
    \plotfiddle{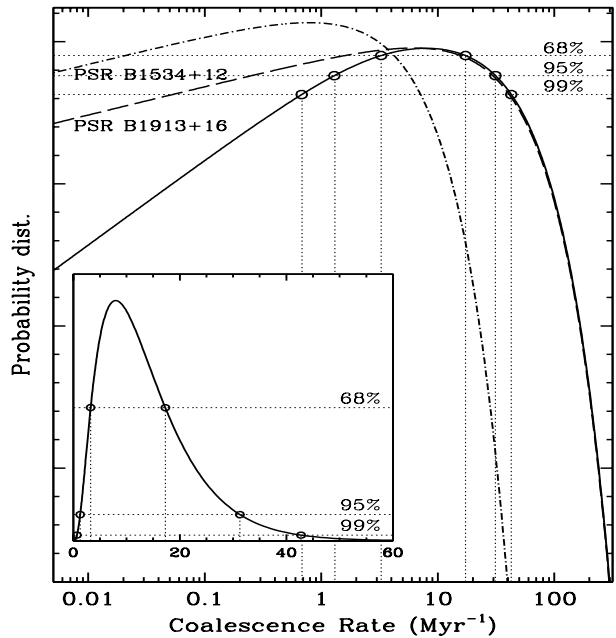}{8.0cm}{0}{43}{33}{-125}{-5}
}
    \caption{The probability distribution function of coalescence rates in
both a logarithmic and a linear scale (small panel). The solid line 
represents $P({\cal R}_{\rm tot})$ and the long and short dashed lines
represent $P({\cal R})$ for PSR~B1913+16-like and PSR~B1534+12-like
populations, respectively.  The dotted lines indicate the confidence levels for
$P({\cal R}_{\rm tot})$. \label{fig:pdf}}
   \end{figure}
%%%

 In Fig.\ 1, $P({\cal R}_{\rm tot})$ along with $P({\cal R}_{\rm 1913})$
and $P({\cal R}_{\rm 1534})$ are shown for the reference model. It is
evident that the total rate distribution is dominated by that of
PSR~B1913+16.  This is due to the fact that we calculate the two rate
contributions having relaxed the constraint that pulsars have luminosities
equal to that of the observed pulsar, and instead allowing for the full
range in luminosity.  In this case any differences in the two separate
rate contributions depend {\em only} on differences in pulse periods, and
widths. Given that the latter are rather small, it makes sense that, for
example, the most likely values of $N_{\rm tot}$ for the two pulsars come
out to be very similar (e.g. $\simeq 390$ and $\simeq 350$, for
PSR~B1913+16 and PSR~B1534+12, respectively, in the reference model).  
Consequently any difference in the rate contributions from the two
populations is due to the difference in lifetimes (about a factor of 10)
for the two observed pulsars (note that the two do not only have similar
$N_{\rm tot}$ estimates, but also similar beaming correction factors).  
Since the lifetime estimate for PSR~B1913+16 is much smaller, the total
rate distribution is dominated by its contribution.

%%% figure2
   \begin{figure}
   \vbox to 6.cm{
   \plotfiddle{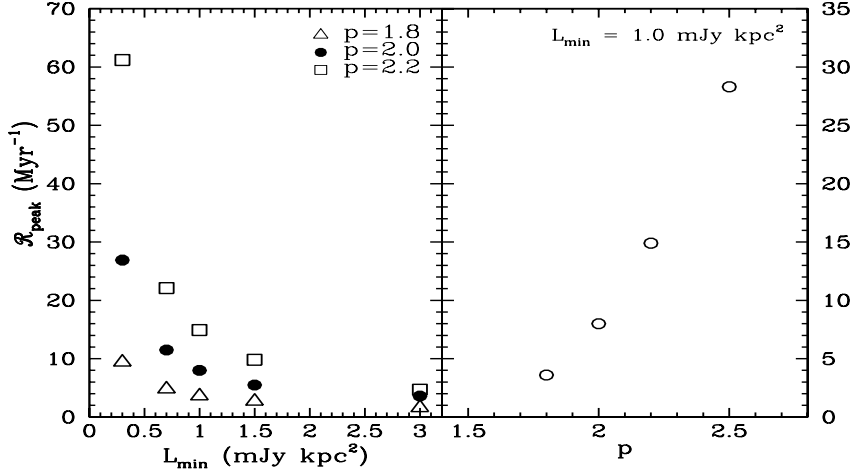}{6.cm}{-90}{42}{32}{-180}{200}
}
   \caption{Left panel: The correlation between ${\cal R}_{\rm peak}$
and the cut-off luminosity $L_{\rm min}$ for different power indices
p of the luminosity distribution function. Right panel: The
correlation between ${\cal R}_{\rm peak}$ and the power index of the
luminosity distribution function $p$. \label{fig:corr1}}
   \end{figure}

We have found that there are strong correlations between the peak
value of the total Galactic coalescence rate, ${\cal R}_{\rm peak}$,
and the cut-off luminosity, $L_{\rm min}$, and its power index, $p$.  
As seen in the Fig.\ 2, ${\cal R}_{\rm peak}$ increases rapidly with
decreasing $L_{\rm min}$ (left panel) or with increasing p (right
panel). The peak rate does not show any strong dependence on scale
lengths of the spatial distribution (either $R_0$ or $Z_0$), except
for rather extreme cases (e.g. $R_{\rm 0} \le 3$ kpc).Hence, the most
important model parameter seems to be the slope and low-end cut-off
of the luminosity function. We find that the most likely values for
the rates are in the range $\simeq$ 3-22\,Myr$^{-1}$ for all models
with luminosity-function parameters consistent with pulsar
observations at 68\% confidence level (Cordes \& Chernoff 1997).

 \section{Probability Distribution of $N_{\rm obs}$ for Parkes Multibeam
Survey}

\subsection{Statistical Method} 

As mentioned in \S 2.1, because of the long integration time, the
signal-to-noise ratio for the PMB survey is severely reduced by Doppler
smearing due to the pulsars' orbital motion. Acceleration searches in the
reanalysis by Faulkner et al.~(these proceedings) promise to alleviate this 
reduction in the near future. Here we calculate the
probability distribution of the number of pulsars $N_{\rm obs}$ that could
be detected with the PMB survey, assuming that the reduction in flux due
to Doppler smearing is corrected perfectly. Using this distribution P($N_{\rm
obs}$), we calculate the average value of $N_{\rm obs}$, $<N_{\rm
obs}>_{\rm PMB}$, for the PMB survey.

We first calculate $N_{\rm obs}$ (detected by PMB survey) for a range of
total number of pulsars in our model galaxy and determine the slope,
${{\alpha}_{\rm PMB}}$ using eq. (2)(\S 3.1). We then use $P(N_{\rm tot})$
(in practice $P(\lambda)$) derived in \S\,3, to calculate $P(N_{\rm
obs})$, for each type of pulsar:
 \begin{equation}
 P(N_{\rm obs}) = \int P({N_{\rm obs}};{\lambda_{\rm PMB}}) P(\lambda_{\rm
PMB})~ {d{\lambda_{\rm PMB}}},
 \end{equation}
 where $P({N_{\rm obs}};{{\lambda}_{\rm PMB}})$ is given in eq. (1).
Defining $\beta= \frac{\alpha}{{\alpha}_{\rm PMB}}$, P($\lambda_{\rm
PMB}$) can be derived from P($N_{\rm tot}$):
 \begin{equation}
 P({\lambda}_{\rm PMB}) = \beta^{2} {{\lambda}_{\rm PMB}} e^{-\beta
{\lambda_{\rm PMB}}}.
 \end{equation}
 Since $\alpha$'s are different for each type of pulsar, $\beta$ is
determined for each pulsar, separately. The normalized probability
distribution of $N_{\rm obs}$ for each type of pulsar, P($N_{\rm obs}$),
is then calculated as: 
 \begin{equation}
 P(N_{\rm obs})={\frac{{\beta}^{2}}{(1+\beta)^{2}}} { \frac{({N_{\rm
obs}}+1)}{(1+{\beta})^{N_{\rm obs}}} },
 \end{equation}

 Once we have P($N_{\rm obs}$)$_{\rm 1913}$ and P(N$_{\rm obs}$)$_{\rm
1534}$, we can calculate the combined $P(N_{\rm obs,1913}+N_{\rm
obs,1534})$.  We define ${N_{\rm +}} = {N_{\rm obs,1913}} + {N_{\rm
obs,1534}}$ and ${N_{\rm -}} = {N_{\rm obs,1913}} - {N_{\rm obs,1534}}$,
where all variables are integers.  The joint probability distribution
function for observing either PSR~B1913+16-like or PSR~B1534+12-like
pulsar, P(${N_{\rm +}}$,${N_{\rm -}}$) is given by:
 \begin{equation}
 P({N_{\rm +}},{N_{\rm -}})= {\Bigl[ {\frac { {\beta_{\rm 1}} {\beta_{\rm
2}} }{ 2({1+{\beta_{\rm 1}}})({1+{\beta_{\rm 2}}}) }}\Bigr]^{2}} {\frac {
({ {N_{\rm +}} + {N_{\rm -}}}+2)({ {N_{\rm +}} - {N_{\rm -}}}+2) } {
({1+{\beta_{\rm 1}}})^{({{N_{\rm +}} + {N_{\rm -}}})/2} ({1+{\beta_{\rm
2}})^{({{N_{\rm +}} - {N_{\rm -}}})/2}} }}.
 \end{equation}
 Finally, we calculate $P(N_{\rm +})$ by the summation of P(${N_{\rm
+}}$,${N_{\rm -}}$) over $N_{\rm -}$.
 \begin{equation}
 P({N_{\rm +}})= \sum _{\rm {N_{\rm -}}} P({N_{\rm +}},{N_{\rm -}}), 
 \end{equation}
where $N_{\rm -}$ lies in the range [$-N_{\rm +}$,$N_{\rm +}$] with an
increment of 2.

\subsection{Results} 

%%% figure3 (pmb.eps)
   \begin{figure}[t]
   \vbox to 6.5cm {
   \plotfiddle{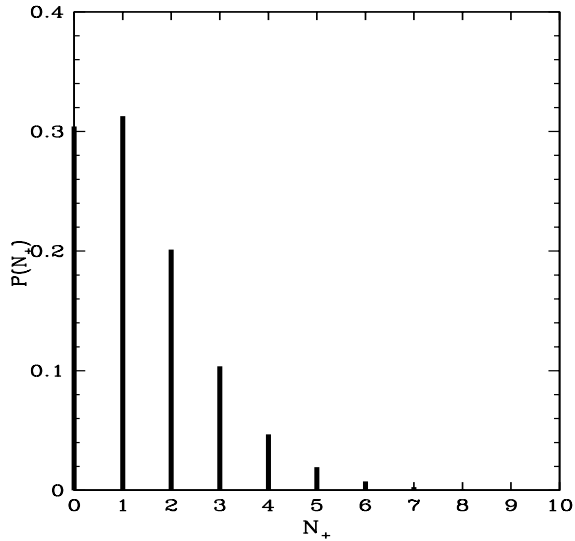}{6.5cm}{0}{38}{28}{-115}{-5}
}
   \caption{The probability distribution function P($N_{\rm obs}$)  for
the PMB survey in the absence of any signal-to-noise ratio due to orbital
motions. Results are shown for our reference model, but they are highly
insensitive to the model parameters. \label{fig:fvst} }
   \end{figure}   
%%%

In Fig.\ 3, we show the distribution function P(${N_{\rm +}}$)  that
could be detected by the PMB survey for our reference pulsar
population model. We have calculated the distributions and average
predicted values, for other models with different cut-off luminosity
$L_{\rm min}$ and the power index of the luminosity function, $p$
(KKL). For 7 different models with luminosity-function parameters
within a 68\% confidence level (Cordes \& Chernoff 1997), we find
values of $<N_+>$ in the range 1.35--1.5 (1.4 for our reference
model). We conclude that, if orbital motion does not affect the
signal-to-noise ratio, then the PMB survey could be expected to
detect 1-2 new binary pulsars with pulse profile and orbital
properties similar to either PSR B1913+16 or PSR B1534+12 in the PMB
survey.

 \section{Discussion}

 We have recently developed a new method for estimating the total number
of pulsars in our Galaxy and have applied it to the calculation of the
coalescence rate of double neutron star systems in the Galactic field (for
more details see KKL). Here, we extend this method to obtain a prediction
for the average number of observed pulsars that the PMB survey could
detect when acceleration searches are used to correct for the Doppler
smearing due to orbital motions.  The modeling of pulsar survey selection
effects is formulated in a ``forward'' way, by populating the Galaxy with
model pulsar populations and calculating the likelihood of the real
observed sample. This is in contrast to the ``inverse'' way of the
calculation of scale factors used in previous studies.

We note that this method could be further extended to account for
distributions of pulsar populations in pulse periods, widths, and
orbital periods. It is important to note that both our rate estimates
and the predictions for detections from the PMB survey do not apply
to binary pulsars that are significantly different from with such
properties that are significantly different PSRs~B1913+16 and
B1534+12 in terms of pulse shapes and orbital properties.

Most importantly the method can be applied to any type of pulsar
population with appropriate modifications of the modeling of survey
selection effects. Currently we are working on assessing the
contribution of double neutron stars formed in globular clusters as
well as the formation rate of binary pulsars with white dwarf
companions that are important for gravitational-wave detection by
LISA, the space-based interferometer planned by NASA and ESA for the
end of this decade.
 
 \acknowledgments
 This work is partially supported by NSF grant PHY-0121420 and a Packard
Science and Engineering Fellowship to VK.  DRL is a University Research
Fellow funded by the Royal Society. DRL is also grateful for the
hospitality and support of the Theoretical Astrophysics Group at
Northwestern U.
 
%\begin{thebibliography}{99}

\end{document}